\documentclass[psfig,letter]{aastex}

\usepackage{emulateapj5}

\def\ltsima{$\; \buildrel < \over \sim \;$}

\def\lsim{\lower.5ex\hbox{\ltsima}}
\def\gtsima{$\; \buildrel > \over \sim \;$}
\def\gsim{\lower.5ex\hbox{\gtsima}}

\begin{document}

\title{Was the Universe Reionized by Massive Metal Free Stars?}

\author{J. Stuart B. Wyithe\altaffilmark{1} and Abraham
Loeb\altaffilmark{2}}

\email{swyithe@isis.ph.unimelb.edu.au; loeb@sns.ias.edu}

\altaffiltext{1}{The University of Melbourne, Parkville, Vic, Australia}

\altaffiltext{2}{Institute for Advanced Study, Princeton, NJ 08540; 
Guggenheim Fellow; On leave from the Astronomy Department, Harvard University}

\begin{abstract}
\noindent 

The {\em WMAP} satellite has measured a large optical depth to
electron scattering after cosmological recombination of $\tau_{\rm
es}=0.17\pm 0.04$, implying significant reionization of the primordial
gas only $\sim 200$ million years after the big bang. 
However, the most recent overlap of intergalactic HII regions must
have occurred at $z\la 9$ based on the Ly$\alpha$ forest constraint on
the thermal history of the intergalactic medium.
Here we argue that a first generation of metal--free stars with a heavy
(rather than Salpeter) mass function is therefore {\em required} to
account for much of the inferred optical depth.
This conclusion holds if feedback regulates star formation in
early dwarf galaxies as observed in present-day dwarfs.

\end{abstract}


\section{Introduction}

The first stars in the universe (Pop-III) formed out of metal-free
gas, relic from the big bang. The bottom-up hierarchy of structure
formation in Cold Dark Matter (CDM) models implies that stars started
to form in CDM halos just above the cosmological Jeans mass of $\sim
10^5M_\odot$ at redshifts as high as $z\sim 20-30$ (see review by
Barkana \& Loeb 2001 and references therein). Simulations of
metal-free star formation indicate that the first stars had high
masses ($M\ga 100M_\odot$) since gas cooling by molecular hydrogen
(H$_2$) cannot lower the gas temperature below $\sim 200$K (Bromm,
Coppi, \& Larson 2002; Abel, Bryan \& Norman 2002).  Massive,
metal-free stars shine at their Eddington luminosity, $L_{\rm E}
\propto M$, and have roughly constant effective (surface) temperatures
of $\sim 10^5$K and lifetimes of $\sim 3\times 10^6$yr, independent of
their mass (Bromm, Kudritzki, \& Loeb 2001). This implies that the
number of ionizing photons produced per baryon incorporated into these
stars ($\sim 4.5\times10^4$) was independent of their mass function,
and larger by more than an order of magnitude than the yield of the
observed Pop-II (low-mass, metal-rich) stars.

{\it Did the first generation of metal-free stars produce a sufficient
total number of ionizing photons to reionize the universe?}  This
question is difficult to answer theoretically.  The UV radiation from
the massive stars can easily dissociate the fragile intergalactic
H$_2$ molecules which are responsible for star formation in mini-halos
with virial temperatures $T_{\rm vir}\la10^{2.4}$K (Haiman, Rees, \&
Loeb 1997; Haiman, Abel \& Rees~2000; Ricotti et al. 2002). Moreover
they could evaporate the gas out of the shallow potential wells of the
first CDM mini-halos (Barkana \& Loeb 1999; Nishi \&
Tashiro~2000). Finally the hydrodynamic energy released by a single
supernova explosion is much larger than the gravitational binding
energy of the gas in these halos (Barkana \& Loeb 2001). If feedback
quenches star formation in mini-halos, then Pop-III star formation
will be dominated by more massive halos where H$_2$ can form in dense
gas shielded from the UV background (Oh \& Haiman~2002; Mackey, Bromm
\& Hernquist~2002).

Analysis of the first year data from the {\em WMAP} satellite suggests a
large optical depth to electron scattering $\tau_{\rm es}=0.17\pm
0.04$, implying significant reionization at
redshifts as high as $\sim 17\pm 5$ (Kogut et al.~2003; Spergel et
al.~2003). While this result is not surprising in view of earlier
detailed calculations of the reionization history by metal-free stars
(Wyithe \& Loeb 2003; Paper I), it offers an intriguing
{\it empirical} path for answering the aforementioned question.  In
this {\it Letter}, we argue that metal-free stars are {\it required} by
the {\em WMAP} measurement of $\tau_{es}$.

We assume the most
recent cosmological parameters obtained through fits to {\em WMAP} data
(Spergel et al.~2003). These include density parameters of
$\Omega_{m}=0.27$ in matter, $\Omega_{b}=0.044$ in baryons,
$\Omega_\Lambda=0.73$ in a cosmological constant, a Hubble constant of
$H_0=71~{\rm km\,s^{-1}\,Mpc^{-1}}$, an rms amplitude of
$\sigma_8=0.84$ for mass density fluctuations in a sphere of radius
$8h^{-1}$Mpc,
and a primordial power-spectrum with a power-law 
index $n=1$. Any tilt in the power spectrum at the level that is
marginally indicated by the combination of the {\em WMAP} data with other
data sets (Spergel et al.~2003) would reduce the small-scale power,
delay reionization, and strengthen our conclusion.
 
\section{Optical Depth to Electron Scattering}
\label{opticaldepth}

Different reionization histories result in different large-scale
polarization anisotropies of the microwave background (Zaldarriaga
1997; Kaplinghat et al. 2002).  A simple measure of the reionization
history is the integrated optical depth to electron scattering
$\tau_{\rm es}=\int_0^{10^3}dz\vert{cdt}/{dz}\vert \sigma_{T} n_{\rm
e}(z)$, where $\sigma_{T}$ is the Thomson cross-section and
$(dt/dz)^{-1}=-H_0(1+z)\sqrt{\Omega_m(1+z)^3+\Omega_\Lambda}$. The
electron number density, $n_{\rm e}$, depends on the mass filling factors
$Q^{\rm m}$ of the different ionized species in the gas,
\begin{equation}
n_{\rm e}=(1+z)^3\left[Q^{\rm m}_{\rm
H^+}n_{\rm H}^{\rm 0} + Q^{\rm m}_{\rm He^+}n_{\rm He}^{\rm
0} + 2Q^{\rm m}_{\rm He^{++}}n_{\rm He}^{\rm 0}\right],
\end{equation}
where $n^0$ denotes the total present-day number densities of hydrogen
(H), and helium (He).  

\section{Reionization Histories}
\label{reionhist}

We now consider reionization histories under different model
assumptions.  The histories follow the reionization of hydrogen as
well as the single and double reionization of helium in an
inhomogeneous IGM (Miralda-Escude et al.~2000). We include helium in
our calculations since the harder spectrum of massive Pop-III stars
can reionize both hydrogen and helium at high redshift (Oh et
al. 2001; Venkatesan, Tumlinson \& Shull~2003).

We calculate the emissivity of quasars described by the luminosity
function of Wyithe \& Loeb (2002a) as well as the emissivity of high
redshift zero-metallicity stars and Pop-II stars (the latter is
discussed in \S~\ref{stars}).  Our model has two free parameters: (i)
the transition redshift, $z_{\rm tran}$, at which Pop-II stars start
to dominate the cosmic radiation field of ionizing photons; and (ii)
the product of the star formation efficiency in massive galaxies
$f_\star$ (=stellar mass/total initial baryonic mass) and the escape
fraction $f_{\rm esc}$ of ionizing photons from galaxies. In Paper I
we presented contours of $\tau_{\rm es}$ as a function of $z_{\rm
tran}$ and $f_\star f_{\rm esc}$. We have found that if the star
formation efficiency in early low-mass galaxies resembled that in
local dwarf galaxies with the same circular velocity, then values of
$f_\star f_{\rm esc}\sim0.05-0.2$ and $z_{\rm tran}\sim15$ lead to
$\tau_{\rm es}$ within the range measured by {\em WMAP}. Below we
summarize the basic ingredients of our model; more details can be
found in Paper I.

We treat the IGM as a two-phase medium. In regions of reionized IGM,
star formation can only be initiated in halos that collapse above a
virial temperature threshold of $T_{\rm min}=2.5\times 10^5$K,
corresponding to the cosmological Jeans mass of photo-ionized IGM at
$\sim10^4$K (e.g., Thoul \& Weinberg~1996; Kitayama \& Ikeuchi~2000).  
In the neutral IGM, star formation is
limited to halos with virial temperatures above the minimum for
efficient gas cooling. If only atomic hydrogen is present, then
$T_{\rm min}=10^4$K. However if molecular hydrogen is present then
$T_{\rm min}=10^{2.4}$K (Haiman, Thoul, \& Loeb 1996; Haiman, Abel, \&
Rees 2000) and star formation can proceed in much smaller halos. We
begin by discussing the case with $T_{\rm min}=10^4$K.

\subsection{Ionizing Photons from Stars}
\label{stars}

To compute the comoving densities of stellar ionizing photons
(${n_{\rm \gamma}}$) we integrate over model spectra for stellar
populations to find the total number of ionizing photons ($N_{\rm
\gamma}$) emitted per baryon incorporated into stars. We find
\begin{equation}
\label{star_ionize}
\frac{dn_{\rm \gamma}}{dz} = N_{\rm \gamma}f_{\rm
esc}\frac{dF_\star}{dz}n_{\rm b},
\end{equation}
where $n_{\rm b}$ is the number density of baryons, and $F_\star(z)$
is the fraction of baryons in the universe at redshift $z$ that
collapsed, cooled, and formed stars.

A recent analysis of SDSS data (Kauffmann et al.~2002) has shown that
in a large sample of local galaxies, the ratio between the total
stellar ($M_\star$) and dark matter halo ($M$) 
masses scales as $M^{2/3}$ for
$M_{\star}<3\times10^{10}M_{\odot}$, but is constant for larger
stellar masses. Thus the star formation efficiency within dwarf galaxies
is proportional to $M^{2/3}$.  Since star formation is thought to be
regulated by supernova feedback (Dekel \& Silk 1986; Dekel \& Woo 2002),
and the supernova energy output is proportional to $M_\star$ while the
gravitational binding energy scales as $M v_{\rm cir}^2 \propto
M^{5/3}$, more massive galaxies retain a larger fraction of their gas.
Using the stellar mass Tully-Fisher relation of Bell \& De
Jong~(2001), we find the threshold circular velocity $v_\star=176~{\rm
km~s^{-1}}$ that corresponds to a stellar mass of
$3\times10^{10}M_{\odot}$ at $z=0$. The mass of a dark matter halo
with circular velocity $v_\star$ is $M^\star_{\rm
halo}\sim5\times10^{10}\left[(1+z)/{10}\right]^{-3/2}M_\odot$.  We
define $f_\star$ as the star formation efficiency in galaxies with
circular velocities larger than $v_\star$; below $v_\star$ the star
formation efficiency is $\eta_\star = f_\star\left({M}/{M_{\rm
halo}^\star}\right)^{2/3}$.  Given $T_{\rm min}$ we find
\begin{equation}
\label{ecolf}
\frac{dF_\star(T_{\rm
min})}{dz}=\frac{d}{dz}\frac{1}{\rho_{\rm m}}\int_{M(T_{\rm
min})}^\infty dM \epsilon f_\star M\frac{dn_{\rm st}}{dM},
\end{equation}
where $\epsilon=1$ for $M>M_{\rm halo}^\star$ and
$\epsilon=\left({M}/{M_{\rm halo}^\star}\right)^{2/3}$ for $M<M_{\rm
halo}^\star$, $\rho_{\rm m}$ is the mean comoving mass density of
matter, and $n_{\rm st}(>M)$ is the comoving number density of halos
with masses $>M$ (Sheth \& Thormen 1999).  
Finally, the value of $dF_\star/dz$ in the two phase
IGM required for calculation of $dn_{\rm \gamma}/dz$ may then be
computed from
\begin{equation}
\label{fb}
\frac{dF_\star}{dz}=Q^{\rm m}_{\rm
H^+}\frac{dF_\star(2.5\times10^5\mbox{K})}{dz} + (1-Q^{\rm m}_{\rm
H^+})\frac{dF_\star(10^4\mbox{K})}{dz}.
\end{equation}

\subsection{The Need for Metal-Free stars}

We begin with calculations of reionization histories in the absence of
metal-free stars at all redshifts.  We have calculated reionization
histories for several values of $f_\star f_{\rm esc}$ in terms of the
volume filling factor $Q$ of ionized regions in the low overdensity
($\la 20$) IGM (Miralda-Escude et al. 2000; Paper I), and
computed the resulting values of $\tau_{\rm es}$.  
  
Recently, Theuns et al. (2002) have argued that if HI reionization
ended before $z=9$, and the universe remained reionized after that,
then adiabatic cooling would have lead to an IGM temperature at
$z=3.5$, much lower than inferred from the Ly$\alpha$ forest. As a
result, the most recent overlap of intergalactic HII regions must have
occurred at $z\la9$, unless there was significant photo-heating at 
$z\gg3.5$ due to the double reionization of helium. In the absence 
of metal free stars we find that
this constraint requires $f_\star f_{\rm esc}\la0.1$, resulting in a
value of $\tau_{\rm es}\la0.09$ that is not consistent with the {\em
WMAP} data. On the other hand, if we neglect the constraint that HII
overlap occurred at $z\la9$, we find that in the absence of metal-free
stars, reionization histories resulting in $\tau_{\rm es}\ga0.13$
imply HII overlap redshifts of $\sim15$.  However assuming feedback
regulation of star formation in low mass galaxies, these histories
require the unphysical value of $f_\star f_{\rm esc}=1.0$ in galaxies
with $v_{\rm cir}>v_\star$.  We therefore conclude that reionization
by an early population of massive metal-free stars (with very
large\footnote{We conservatively adopt the ionizing photon yield of
$~100M_\odot$ metal-free stars. The yield is larger by a factor of
$\sim 2$ for Pop-III stars with masses $\ga300M_\odot$ (Bromm,
Kudritzki, \& Loeb 2001). The use of the higher yield would have lead
to the same $\tau_{\rm es}$ for values of $f_\star f_{\rm esc}$ at
$z>z_{\rm tran}$ that are smaller by a factor of 2 than we quote.}
$N_\gamma$) must have contributed significantly to $\tau_{\rm es}$ at
$z\sim 15$.

The formation of metal free stars is moderated with cosmic time by the
gradual metal enrichment of the IGM. It has been argued that the
dominant mode of star formation change to Pop-II once the metallicity
of the associated gas rises above $10^{-4}$Z$_\odot$ (Schneider et
al. 2002; Bromm et al. 2001). If all Pop-III stars formed in the mass
range of $140$M$_\odot-260$M$_\odot$ where pair-unstable supernovae
result in expulsion of metals into the surrounding gas (Heger \&
Woosley~2001), and if all metals were mixed uniformly throughout the
entire IGM, then Pop-III star formation would have ceased as soon as
$\sim0.3$ ionizing photons were produced per cosmic hydrogen atom
(Ciardi, Ferrara \& White~2003). However, the highly inhomogeneous
nature of the enrichment process (Furlanetto \& Loeb 2003) and the
formation of Pop-III stars beyond the above mass range, naturally
lead to reionizion.

\begin{figure*}[htbp]
\epsscale{1.9}
\plotone{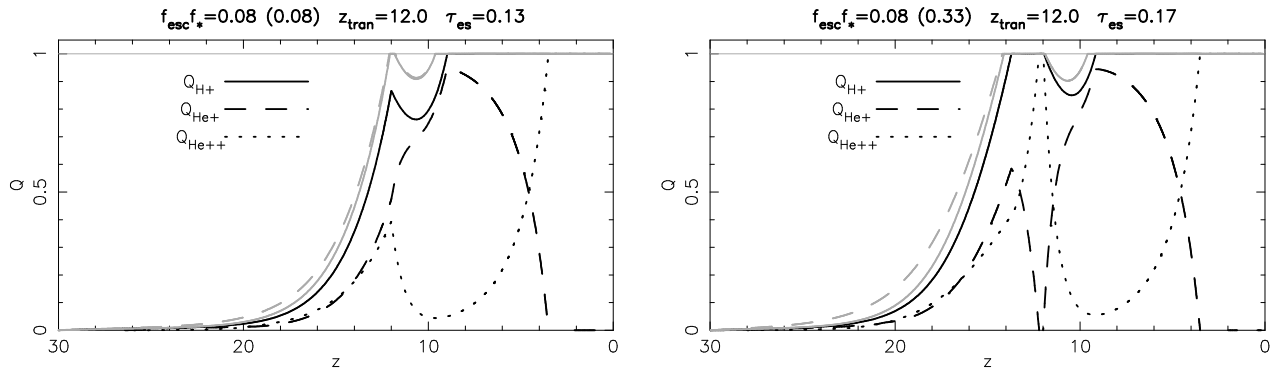}
\caption{\label{plot1} Sample reionization histories
assuming feedback regulated star formation, that produce large values
of $\tau_{\rm es}$. Dark lines represent the evolution of the volume
filling factors of ionized regions $Q_{\rm H^+}$, $Q_{\rm He^+}$ and
$Q_{\rm He^{++}}$. We also show calculations of $Q_{\rm H^+}$ in the
absence of helium (light lines). The solid light lines show results
for a pre-reionization minimum virial temperature for galaxies of
$T_{\rm min}=10^4$K (atomic cooling) while the dashed light line
corresponds to $T_{\rm min}=10^{2.4}$K (molecular cooling). 
Each example is labeled by the values of $f_\star f_{\rm
esc}$ at $z<z_{\rm tran}$ (and in parenthesis at $z>z_{\rm tran}$),
$z_{\rm tran}$ and $\tau_{\rm es}$.}
\end{figure*}

{\it Must the Pop-III stars possess a heavy initial mass function
(IMF)?} Tumlinson \& Shull~(2000) computed the ionizing photon output
per baryon of zero-metallicity stars with a Salpeter IMF and found
only a 50\% increase over the values for metal enriched Pop-II stars
with the same IMF. This moderate increase is to be compared with the
increase by more than an order of magnitude for the massive Pop-III
stars described by Bromm, Kudritzki, \& Loeb (2001) (the number of ionizing
photons produced per baryon is nearly independent of stellar mass).  
The arguments presented above therefore imply that Pop-III stars were very 
massive rather than of a Salpeter IMF.
This provides the first observational support to the results from
hydrodynamic simulations of early star formation (Bromm et al. 2002;
Abel et al.  2002).

In generating the star formation history we have extrapolated results
from the local galaxy sample, assuming that feedback governs the
observed stellar mass as a function of halo potential depth.  However
high redshift galaxies may have cooling times that are shorter than the
dynamical time of their cold gas. This may lead to a starburst that
might render the time-delayed supernovae feedback less effective. We now
briefly consider this case. Assuming the star formation efficiency to
be independent of halo mass (case A in Paper I), we find that
reasonable values of $f_\star f_{\rm esc}\sim 0.1$ can yield
$\tau_{\rm es}\sim0.17$ in the absence of massive Pop-III
stars. However, the corresponding reionization histories also have an
early HII overlap. Therefore, unless the global product of escape
fraction and star formation efficiency was more than an order of
magnitude larger at $z\sim15$ than at $z\sim9$, our conclusion that an
early generation of massive metal free stars is required to produce
both $\tau_{\rm es}\sim0.17$ and HII overlap at $z\la9$ is independent
of whether feedback regulates star formation in low mass galaxies or
not.

\subsection{Reionization Histories}
\label{rhs}

We now show that the joint constraints of $\tau_{\rm es}=0.17\pm0.04$
and HII overlap at $z\la9$ can be met simultaneously for values of
$z_{\rm tran}>9$ and $f_\star f_{\rm esc}<1$ if massive metal-free
stars were present at high redshifts.  In
Figure~\ref{plot1} we show two cases of the evolution of $Q$ for
hydrogen (solid lines), singly ionized helium (dashed lines) and
doubly ionized helium (dotted lines).  We assume\footnote{Salvaterra
\& Ferrara~(2003) have recently presented the observational case that
$z_{\rm tran}\la9$ through fitting the near-infrared background.  The
lower value of $z_{\rm tran}$ would allow a
second HII overlap near $z\sim7$ to be compatible with the large value
of $\tau_{\rm es}$ and IGM temperature constraints.} 
$z_{\rm tran}=12$ and $f_\star f_{\rm esc}=0.08$ at
$z<z_{\rm tran}$, yielding the most recent HII overlap at $z\sim9$.
At $z>z_{\rm tran}$ we assume $f_\star f_{\rm esc}=0.08$ (left) and
$f_\star f_{\rm esc}=0.33$ (right). The latter allows for the
possibility of a higher star formation efficiency\footnote{Note that
the star formation efficiency in the low-mass galaxies of interest,
$\eta_\star= 0.02
f_\star\left(M/10^{8}M_\odot\right)^{2/3}[(1+z)/15]$, is generally
$\ll f_\star$.  }  or escape fraction (Steidel et
al. 2001) at high redshift. In the former case there is a significant
reionization but {\it failed} HII overlap at early times, followed by
a recombination and a successful HII overlap at $z\sim9$ due to Pop-II
stars. The two examples yield $\tau_{\rm es}$ within one standard
deviation and at the most likely value of {\em WMAP}, respectively.
Note that our model leads to the double reionization of helium by
quasars at $z\sim3.5$ independent of the contribution from Pop-III
stars.  This is in excellent agreement with current observational data
[e.g. Theuns et al.~(2002) and references therein].  The reionization
of HeIII by quasars is characterized by a sharp rise in our model,
with $Q_{\rm He^{++}}$ increasing from $\sim50\%$ at $z\sim4.3$ to
$\sim 100\%$ at $z\sim 3.7$. We do not expect pre-heating
from the reionization of HeIII to impact on the conclusion
that the most recent HII overlap took place at $z\la9$.

Figure~\ref{plot1} also shows the evolution of
$Q$ for hydrogen when helium is ignored (light solid line). In these
calculations all photons blueward of 13.6eV reionize hydrogen. We find
that the additional ionizing photons result in overlap of hydrogen at
an earlier epoch, but the net optical depth to electron scattering is
decreased.

So far we have assumed that atomic cooling dictates the mass of the
smallest galaxies. However it has been suggested that significant
production of X-rays could catalyze molecular hydrogen production
allowing star formation in mini-halos before reionization (Haiman,
Rees, \& Loeb 1996; Cen 2002). The light dashed lines in the left hand
panels of Figure~\ref{plot1} show the evolution in $Q$ for hydrogen (in
the absence \ of helium) assuming $T_{\rm min}=10^{2.4}$K in the
neutral IGM. We find that the feedback-regulation built into the model
sufficiently limits the star formation efficiency in mini-halos so
that they do not contribute significantly to the reionization history.

\section{Conclusions}
\label{conc}

We have shown that Pop-II stars alone cannot reionize the universe
sufficiently early to produce the value of $\tau_{\rm es}\sim
17\pm4\%$ inferred from the {\em WMAP} data (Kogut et al.~2003;
Spergel et al.~2003).  This conclusion holds if feedback regulates
star formation in early low-mass galaxies as observed in present-day
dwarf galaxies. Without metal free stars, the condition that the most
recent overlap of intergalactic HII regions must have occurred at
$z\la 9$ (Theuns et al. 2002) generically yields $\tau_{\rm es}\la
9\%$. The joint requirements that $\tau_{\rm es}\sim 17\%$ and that
HII regions overlapped at $z\la9$ imply that metal-free stars
reionized a significant fraction of the IGM at $z\sim15$.  Moreover,
we find that these stars must have been {\it massive} to provide the
required order-of-magnitude increase in the production rate of
ionizing photons relative to low mass Pop-II stars (Bromm, Kudritzki \&
Loeb~2001).  Pop-III stars with a Salpeter IMF provide only a $50\%$ 
increase in $N_\gamma$ (Tumlinson \& Shull 2000) and cannot
account for the large $\tau_{\rm es}$.

For plausible reionization histories with metal-free stars we have
found that singly and doubly ionized helium contributes as much as
$\Delta \tau_{\rm es}=0.02$, and should be included in reionization
calculations.  We also found that if the star formation efficiency is
regulated in dwarf galaxies as observed in the local universe, then
there is no significant contribution to the cosmic ionizing radiation
field from halos with virial temperatures $T_{\rm vir}\la10^4$K, even
if molecular hydrogen permits cooling of gas inside these mini-halos.

The evolution of the ionization state and metal enrichment of the IGM
during the reionization epoch can be best probed by infrared
spectroscopy of gamma-ray burst (GRB) afterglows (Furlanetto \& Loeb
2002).  These transient sources should be both abundant (Bromm \& Loeb
2002) and detectable (Lamb \& Reichart~2000; Ciardi \& Loeb 2000) out
to high redshifts.  The {\it Swift} satellite, planned for launch this
year, is expected to detect $\sim 100$ GRBs per year, out of which
$\ga 20\%$ may originate at $z\ga 5$ (Bromm \& Loeb 2002).
Photometric identification of the Ly$\alpha$ trough 
can be used to select the high-redshift afterglows for which
spectroscopic follow-up observations may be conducted on a 10-m class
telescope (Loeb 2001).

After the submission of this paper, a number of different groups have
also commented on the implications of the large $\tau_{\rm es}$
measured by {\em WMAP}. Some studies find that Pop-II stars are able
to reionize the universe sufficiently early to produce $\tau_{\rm
es}\ga0.13$ (Ciardi, Ferrara \& White~2003; Somerville \&
Livio~2003); however, this conclusion is challenged by the existence
of the observed Gunn-Peterson trough in quasars at $z\sim6.3$ (as
noted by Ciardi et al.~2003) and by the temperature evolution of
the IGM (Theuns et al. 2002; Hui \& Haiman~2003).  Other authors agree
that the combined requirements of a large $\tau_{\rm es}$ and a late
HII overlap implies that the efficiency of production of ionizing
photons was higher at redshifts $z\ga 10$ (Haiman \& Holder~2003;
Sokasian, Abel, Hernquist \& Springel~2003; Cen~2003). The conclusion
that Pop-III stars needed to be massive to produce the observed
$\tau_{\rm es}$ was also reached by Cen~(2003) in agreement with our
results.

\acknowledgements 

We thank Rennan Barkana, Volker Bromm, Steve Furlanetto, Rachel
Webster and the referee Andrea Ferrara for very useful comments on the
manuscript.  AL acknowledges generous support from the Institute for
Advanced Study at Princeton and the John Simon Guggenheim Memorial
Fellowship.  This work was also supported in part by NSF grants
AST-0071019, AST-0204514 and NASA grant ATP02-0004-0093.


\begin{thebibliography}{}

\bibitem[]{485}
Abel, T., Bryan, G. L., \& Norman, M. L. 2002, Science, 295, 93

\bibitem[]{488} 
Barkana, R.~\& Loeb, A.\ 1999, \apj, 523, 54

\bibitem[]{491} 
Barkana, R.~\& Loeb, A.\ 2001, Physics Reports, 349, 125

\bibitem[Bell \& de Jong(2001)]{2001ApJ...550..212B} 
Bell, E.~F.~\& de Jong, R.~S.\ 2001, \apj, 550, 212

\bibitem[]{BCL2002}
Bromm, V., Coppi, P. S., \& Larson, R. B. 2002, ApJ, 564, 23

\bibitem[]{BKL2001}
Bromm, V., Kudritzki, R. P., \& Loeb, A. 2001, ApJ, 552, 464

\bibitem[Bromm \& Loeb(2002)]{2002ApJ...575..111B} 
Bromm, V.~\& Loeb, A.\ 2002, \apj, 575, 111 

\bibitem[]{506}
Cen, R., 2002, astro-ph/0210473

\bibitem[]{509}
Cen, R., 2003, astro-ph/0303236

\bibitem[Ciardi \& Loeb(2000)]{2000ApJ...540..687C} 
Ciardi, B.~\& Loeb, A.\ 2000, \apj, 540, 687 

\bibitem[]{515} 
Ciardi, B., Ferrara, A., White, S.D.M.\ 2003, astro-ph/0302451 

\bibitem[Dekel \& Silk(1986)]{1986ApJ...303...39D} 
Dekel, A.~\& Silk, J.\ 1986, \apj, 303, 39

\bibitem[]{521}
Dekel, A., \& Woo, J. 2002, MNRAS, submitted; astro-ph/0210454

\bibitem[Furlanetto \& Loeb(2002)]{2002ApJ...579....1F} 
Furlanetto, S.~R.~\& Loeb, A.\ 2002, \apj, 579, 1 

\bibitem[]{527} 
Furlanetto, S.~R.~\& Loeb, A.\ 2003, astro-ph/0211496 

\bibitem[Haiman, Rees, \& Loeb(1996)]{1996ApJ...467..522H} 
Haiman, Z., Rees, M.~J., \& Loeb, A.\ 1996, \apj, 467, 522 

\bibitem[Haiman, Rees, \& Loeb(1997)]{1997ApJ...476..458H} 
Haiman, Z., Rees, M.~J., \& Loeb, A.\ 1997, \apj, 476, 458; 484, 985

\bibitem[Haiman, Thoul, \& Loeb(1996)]{1996ApJ...464..523H} 
Haiman, Z., Thoul, A.~A., \& Loeb, A.\ 1996, \apj, 464, 523 

\bibitem[Haiman, Abel, \& Rees(2000)]{2000ApJ...534...11H} 
Haiman, Z., Abel, T., \& Rees, M.~J.\ 2000, \apj, 534, 11 

\bibitem[]{542} 
Haiman, Z., Holder, G., 2003, astro-ph/0302403

\bibitem[Heger \& Woosley(2002)]{2002ApJ...567..532H} 
Heger, A.~\& Woosley, S.~E.\ 2002, \apj, 567, 532 

\bibitem[]{548} 
Hui, L., Haiman, Z., 2003, astro-ph/0302439

\bibitem[]{551} 
Kauffmann, G. et al., 2002, MNRAS accepted, astro-ph/0205070

\bibitem[]{554} Kaplinghat, M., Chu, M., Haiman, Z., Holder, G.~P., Knox,
L., \& Skordis, C.\ 2003, \apj, 583, 24

\bibitem[Kitayama \& Ikeuchi(2000)]{2000ApJ...529..615K} 
Kitayama, T.~\& Ikeuchi, S.\ 2000, \apj, 529, 615 

\bibitem[]{560} 
Kogut, A. et al. 2003, ApJ, submitted; astro-ph/0302213

\bibitem[Lamb \& Reichart(2000)]{2000ApJ...536....1L} 
Lamb, D.~Q.~\& Reichart, D.~E.\ 2000, \apj, 536, 1 

\bibitem[]{566}
Loeb, A. 2001, in ``Supernovae and Gamma-Ray Bursters'', edited by K. W.
Weiler, Springer-Verlag, in press; astro-ph/0106455 

\bibitem[]{570} 
Mackey, J., Bromm, V., Hernquist, L., 2002, astro-ph/0208447

\bibitem[Miralda-Escud{\' e}, Haehnelt, \& Rees(2000)]{2000ApJ...530....1M}
Miralda-Escud{\' e}, J., Haehnelt, M., \& Rees, M.~J.\ 2000, \apj, 530, 1


\bibitem[]{579} 
Nishi, R.~\& Tashiro, M.\ 2000, \apj, 537, 50

\bibitem[]{582} 
Oh, S.~P., Haiman, Z.\ 2002, \apj, 569, 558, 
L1 

\bibitem[Oh, Nollett, Madau, \& Wasserburg(2001)]{2001ApJ...562L...1O} 
Oh, S.~P., Nollett, K.~M., Madau, P., \& Wasserburg, G.~J.\ 2001, \apjl, 562, 
L1 


\bibitem[]{593} 
Ricotti, M., Gnedin, N.~Y., \& Shull, J.~M.\ 2002, \apj, 575, 49

\bibitem[]{596} 
Salvaterra, R., Ferrara, A., 2003, MNRAS, 339, 973

\bibitem[]{599} 
Schneider, R., Ferrara, A., Natarajan, P., Omukai, K.\ 2002, \apj, 571, 30 

\bibitem[Sheth \& Tormen(1999)]{1999MNRAS.308..119S} 
Sheth, R.~K.~\& Tormen, G.\ 1999, \mnras, 308, 119 

\bibitem[]{605} 
Sokasian, A., Abel, T., Hernquist, L., Springel, V., 2003, astro-ph/0303098

\bibitem[]{608} 
Somerville, R.S., Livio, M., 2003, astro-ph/0303017

\bibitem[]{611} 
Spergel, D. et al. 2003, ApJ, submitted; astro-ph/0302209

\bibitem[Steidel, Pettini, \& Adelberger(2001)]{2001ApJ...546..665S} 
Steidel, C.~C., Pettini, M., \& Adelberger, K.~L.\ 2001, \apj, 546, 665 

\bibitem[Theuns et al.(2002)]{2002ApJ...567L.103T} 
Theuns, T., Schaye, J., Zaroubi, S., Kim, T., Tzanavaris, P., 
\& Carswell, B.\ 2002, \apjl, 567, L103 

\bibitem[]{621}
Thoul, A.A. \& Weinberg, D.H., 1996, ApJ, 465, 608

\bibitem[Tumlinson \& Shull(2000)]{2000ApJ...528L..65T} 
Tumlinson, J.~\& Shull, J.~M.\ 2000, \apjl, 528, L65 

\bibitem[]{627}
Venkatesan, A., Tumlinson, J. \& Shull, J.M.\ 2003, \apj, 584, 621


\bibitem[Wyithe \& Loeb(2002)]{2002ApJ...581..886W} 
Wyithe, J.~S.~B.~\& Loeb, A.\ 2002a, \apj, 581, 886 

\bibitem[]{636} 
Wyithe, J. S. B., \& Loeb, A. 2003, ApJ, in press; astro-ph/0209056

\bibitem[]{639} 
Zaldarriaga, M.\ 1997, \prd, 55, 1822



\end{thebibliography}
\end{document}